\documentclass[journal]{IEEEtran}
\usepackage{xcolor,soul,framed} 

\colorlet{shadecolor}{yellow}
\usepackage[pdftex]{graphicx}
\graphicspath{{../pdf/}{../jpeg/}}
\DeclareGraphicsExtensions{.pdf,.jpeg,.png}

\usepackage[cmex10]{amsmath}
\usepackage{array}
\usepackage{mdwmath}
\usepackage{mdwtab}
\usepackage{eqparbox}
\usepackage{url}

\newcommand{\note}[1]{\textcolor{black}{#1}}

\usepackage{subcaption}
\usepackage{subfiles}
\usepackage{float}
\usepackage{graphicx}
\usepackage{xcolor}
\usepackage{amsthm}
\usepackage{amsfonts}
\usepackage{amsmath}
\usepackage{authblk}
\usepackage[switch]{lineno} 

\begin{document}
\bstctlcite{IEEEexample:BSTcontrol}
    \title{Deep Reinforcement Learning for RAN Optimization and Control}


  \author{Yu~Chen, Jie~Chen, Ganesh~Krishnamurthi, Huijing~Yang, Huahui~Wang, Wenjie~Zhao

  \thanks{Y. Chen is with Neuroscience Institute and Machine Learning Department, Carnegie Mellon University, 5000 Forbes Ave. Pittsburgh, PA USA Email: \texttt{yuc2@andrew.cmu.edu}. This work was done when Y. Chen was in a summer internship program at AT\&T Labs . }
  \thanks{Jie~Chen, Ganesh~Krishnamurthi, Huijing~Yang, Huahui~Wang, Wenjie~Zhao are with AT\&T Labs, 1 AT\&T Way, Bedminster, NJ USA. Emails:{\tt \{jc988u, gk4291,hy634t,hw4362,wz5327\}@att.com} } }%


\maketitle

\begin{abstract}
Due to the high variability of the traffic in the radio access network (RAN), fixed network configurations are not flexible enough to achieve optimal performance. Our vendors provide several settings of the eNodeB to optimize the RAN performance, such as media access control scheduler, loading balance, etc. But the detailed mechanisms of the eNodeB configurations are usually very complicated and not disclosed, not to mention the large key performance indicators (KPIs) space needed to be considered. These make constructing a simulator, offline tuning, or rule-based solutions difficult. We aim to build an intelligent controller without strong assumption or domain knowledge about the RAN and can run 24/7 without supervision. To achieve this goal, we first build a closed-loop control testbed RAN in a lab environment with one eNodeB provided by one of the largest wireless vendors and four smartphones. Next, we build a double Q network agent trained with the live feedback of the key performance indicators from the RAN. Our work proved the effectiveness of applying deep reinforcement learning to improve network performance in a real RAN network environment.
\end{abstract}


%
\IEEEpeerreviewmaketitle


\section{Introduction} 
\note{
As the mobile network and the traffic demands grow dramatically over the past years, it becomes critical to proactively optimizing the configurations of the network while minimizing human intervention in the development process. Traditionally, network configurations are static and require frequent fine-tuning to adapt to network dynamics. The optimization cycle could be prohibitively long and costly, as it involves a significant amount of manual work.
}

\note{
There are several challenges for rule-based algorithms to achieve optimal results. First, the development cycle is generally long and involves a significant amount of manual work. It could take months or even years from setting rules to testing and then fine tuning. Human intervention is required during the whole process. Secondly, an optimal RAN control rule/policy may need to coordinate among a large amount of KPIs that can change dynamically. It is difficult for static rules to include all those impacting factors to achieve best results. Therefore rule-based algorithms are not the most desirable solution. Ideally we hope the adjustment could be automatic, agile, and driven by data.
}

\note{
The development of machine learning techniques, especially the emergence of reinforcement learning (RL) \cite{sutton2018reinforcement}, attract a lot of attention in the research community of wireless communications. 
The RL frameworks have been applied on a large variety of network entities, such as the Internet of Things (IoT), Vehicular Ad hoc NETwork (VANET), Dynamic Adaptive Streaming over HTTP (DASH), Long Term Evolution (LTE) networks, 5G networks, etc. 
These works emphasize different aspects of the performance, such as throughput \cite{wang2018deep,liao2019learning,liao2020licensed}, communication latency \cite{ye2018deep}, spectrum cost \cite{challita2017proactive}, cache content \cite{he2017deep}, network bandwidth \cite{chen2017liquid}, the quality of experience \cite{gadaleta2017d,chinchali2018cellular}, etc. \cite{luong2019applications} provide a good survey in these topics. All these studies are based on simulators.
}

\note{
This paper focuses on using deep RL algorithms to automatically select the proper MAC scheduler configuration for each eNodeB. The scheduler allocates the available time-frequency radio resources to users at every transmission time interval \cite{kawser2012performance}. Popular scheduler options include the Round Robin, the Maximum Rate, and the Proportional Fair, etc \cite{kawser2012performance,saxena2016performance,pianese2016optimized}. Different options focus on different objectives, such as equalizing user throughput, maximizing overall spectral efficiency, trading efficiency with fairness, etc. These configuration options are generally provided by the eNodeB vendors to the operators. The selection of scheduling options could significantly impact the throughput of individual users, fairness among users, as well as the capacity of the cell \cite{kawser2012performance}.
}

\note{
Unlike the work in \cite{shoaei2019reconfigurable,al2020learn}, where RL algorithms are applied to directly control resource allocations per individual users in each time slot and resource block group, this paper is working on using the RL algorithm to select the scheduler options by treating the detailed functionality of the scheduler as a black box. The granularity of the control interval can be at minute/hour levels, which is more robust to network dynamics than the prior arts. Because of its generality, the framework implemented in this paper could be easily extended to include other eNodeB configurations and therefore serves as a centralized RAN controller. Another difference distinguishing this work from others is that we are not using simulators for RL training. The limitation of using simulators is obvious: not many simulators can be realistic enough to reflect the schedulers in real networks. Even if there exists a simulator that captures details of the reality, it could take prohibitively long to perform training, because running simulations per se is time-consuming. Additionally, most of such works assume zero delay between the RL agent’s computation and application of the actions to the network, which is not practical, especially in high-frequency control tasks (e.g., a turnaround period of 1ms or even less).
}

\note{
In this paper, we build a closed-loop control testbed in the lab environment, utilizing a real eNodeB controlled by a deep RL agent. We build an interface to allow the agent to automatically change the configuration of the scheduler. A feedback loop is constructed to collect real-time statistics of the Key Performance Indicators (KPIs), which are used as input to the agent to change its actions, i.e., choosing different scheduler options from time to time to optimize the specified reward. The reward can be flexibly defined, adding one more dimension of control for the network operators. By doing this, eNodeB can adaptively adjust its scheduler configuration in response to the changes of the network statistics to achieve optimal outcomes. 
}

The rest of this paper is organized in the following order. Section \ref{sec:experiment} describes each platform module and the experiment paradigm in details. Section \ref{sec:RL_model} gives the details about the RL algorithm. Section \ref{sec:results} presents the results. We extend our ideas and concerns in section \ref{sec:discussion}. Finally, the paper is closed by the conclusion section \ref{sec:conclusion}.

\section{Experiments} \label{sec:experiment}

\begin{figure}[ht]
  \centering
  \includegraphics[width=0.4\textwidth]{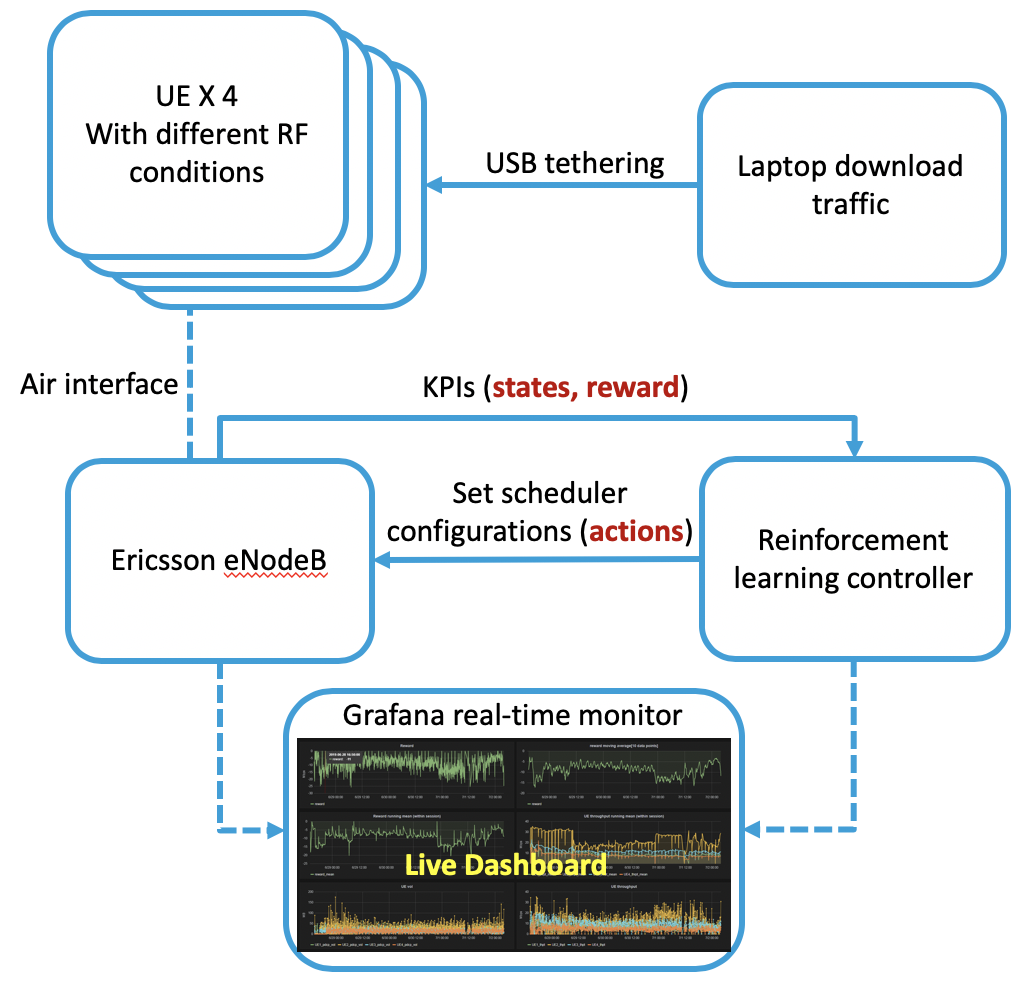}
  \caption{The framework of the RAN platform, which consists of one eNodeB, four UEs and some supporting servers. The reinforcement learning agent controls the configurations of the eNodeB.}
  \label{fig:toycell}
\end{figure}

One of the main contributions of this work is that we set up a reinforcement learning testbed in a RAN test room, which consists of eNodeBs inside and completely blocks outside signals to affect the internal RF environment. We trained and tested the artificial intelligent agent on it. The diagram of the experiment platform is shown in Figure \ref{fig:toycell}. 

To automate the environment and implement live RL, we utilized the API provided by the vendor to change the eNodeB configurations remotely for action change. We collected the rewards/states with Kafka digesting eNodeB's live call trace records. Besides, we used a laptop to control/schedule user equipments' (UEs') behaviors to guarantee test repeatability. Meanwhile, we built a \textit{Grafana} dashboard to monitor the system online and trace history easily. The key components are described in the following paragraphs.

\paragraph{eNodeB}
There are many sets of configurations for an eNodeB, including MAC scheduler, handover offsets, power on/off, transmit power, sector tilt, remote radio heads, baseband units, etc. This work only focuses on one set, the MAC scheduler. Our vendor provides five options for the eNodeB MAC scheduler configuration. Each one corresponds to a different setting for handling resource allocation. The options range from ``equal rate" to ``maximize the UE throughput with the best RF". These options are used to balance the trade-off between throughput and fairness. 
Note that the design of the detailed settings is not provided by the vendor, thus rule-based scheduler configuration or training through simulator are not reliable and impractical due to the extreme complication.

\paragraph{User equipments}
Four Samsung Android smartphones were used as UEs, they were set in USB tethering hotspot mode, and all applications were muted. Then they were tethered with laptops using USB cables, which controlled all the traffics. The upload/download traffic was generated by the laptops using \texttt{iperf} command tool provided by a leading company in wireless. The function is similar to the public \texttt{perf} network performance counter command tool, except that it is specialized for the cellular network owned by the wireless company. It guarantees the traffic goes through the same link path and thus reduces the artifacts introduced by the disturbance of the network. All the traffics from the laptop went through the UEs. To create more realistic traffic, we analyzed the historical traffic logs of eNodeB measurement reports collected from a busy sample cell. This was because the scheduler played a more important role when the cell was handling multiple UEs simultaneously. Usually, an eNodeB might connect to many UEs at the same time within the service area, so we mimic this behavior with 4 UEs in our test. For the selected sample cell and time period, we got each data session's traffic volume and RF condition by joining RRC traffic volume with reference signal received power (RSRP), reference signal received quality (RSRQ) measurement report. The RRC and RF conditions were learned from these recorded data. The k-mean clustering algorithm was performed based on RF conditions and RRC sessions with four clusters. The clusters' centers gave the means of RF conditions and RRCs. We calculated the variance of the RRC using the data from each cluster. Since the time dependency was small, the UEs' RRCs were generated independently from Normal distributions with the estimated means and variances. Also, we adjusted the UEs' RF conditions by changing the positions in the lab or covering them by Faraday bags. The RSRP values were -115dBm, -110dBm, -105dBm, and -94dBm respectively. Standalone tests were conducted to confirm the significant impact of RSRP values on the UE throughput in the lab.

\paragraph{Communication servers, key performance indicators, and monitor server}
In our platform, a KPI composer/publisher server fetched the raw information from the eNodeB, then calculated the KPIs from the raw data and sent them to the reinforcement learning agent. The KPIs of the eNodeB were sampled every minute, such as the spectrum efficiency, cell throughput, number of active UEs, etc. If the sampling frequency was too high, then many data transmission package could not be finished. If the sampling frequency was too low, the status of the network was not well captured.

The KPIs were selected to reflect the quality of the service. We followed the technical specifications (TS 32.450) to calculate the values \cite{3GPPTS}. The KPIs included information on both cell level and UE level. As for cell level KPIs, it included uplink and/or downlink volume/throughput, physical resource block (PRB), control channel element utilization (CCE), neighbor cells relations, handovers, frequency bandwidth, user geographic distribution, reference signal received power (RSRP), reference signal received quality (RSRQ), channel quality information distribution (CQI), timing advance distribution (TA), cell bitrate, cluster harmonic throughput, etc. As for UE level KPIs, it contained video user downlink throughput (video-specific), radio frequency conditions (RF), number of active UEs, UE harmonic throughput, UE throughput gap (the difference between the maximum and minimum UE throughputs), worst UE throughput, etc. These KPIs would be later formatted into states or reward for the reinforcement learning agent. Then the agent updated the states and sent an optimal action to operate the eNodeB.

\paragraph{Reinforcement learning agent}
In the closed-loop training, the RL agent needed the RAN states and the feedback reward as the input, and the actions operating the eNodeB as the outputs. The states were defined as part of the KPIs introduced above. The actions were set as eNodeB MAC scheduler options. The reward was set in two ways; one is the eNodeB overall throughput, the other one is the gap between UEs' throughput, which was used for users' fairness. The details of the RL algorithm will be discussed in section \ref{sec:RL_model}. The reward can be defined in other ways for different purposes. We extend the discussion in section \ref{sec:discussion}.

\paragraph{Experiment paradigm}
In one training episode, the pseudo-random users' demands lasted for 80 minutes, followed by a 10 minutes rest, during which the RL agent was not trained. The demands episodes were repeated so that the evaluation would be easier than comparing the average performance obtained from different situations. This could eliminate the artifacts potentially caused by the users' different demand patterns. \note{
Besides the training episodes, some baseline episodes were arranged by setting the policy with different constant actions. This corresponds to the best outcome of a multi-armed bandit method.
Two sets of experiments were carried out, one was to achieve better cell throughput, 
the baseline policy was the “maximize the UE throughput with the best RF" configuration.
The other one was to minimize the gap of UEs' throughput, the baseline policy was the ``equal rate" configuration. The baseline policies agreed with our expert's optimal design using domain knowledge. 
}

\section{Reinforcement learning model } \label{sec:RL_model}
The radio resources allocated at any moment would affect all the active UEs' sessions. When a session lasts long, it also has an impact on new coming UEs. Therefore, the eNodeB configuration change at any moment has a long-term influence on the network.
Thus, the scheduling process was modeled as a Markov decision process. In general, the goal of the RL is to find an optimal policy that can maximize the discounted sum of the rewards, not just the short-sighted immediate reward. In this way, the delayed reward can be taken into consideration. This property makes RL attractive in many control problems. Finding such an optimal policy is difficult, especially with the high dimensional state space. The recent development of RL has made exciting progress by taking advantage of the deep neural network over Q-learning \cite{mnih2015human}. 

We define the \textit{state} $S_t$ at time $t$ using the KPIs received from the eNodeB, a vector of 58 entries. It contains the KPI values such as the spectrum efficiency, cell throughput, proportion of the utility, channel quality indicator (CQI) ratio, etc. The state does not include the information about individual UEs, as the framework is designed for general purpose, not relying on specific UEs.
The \textit{action} $A_t$ is discrete, which contains all possible eNodeB MAC scheduler configuration options \texttt{EQUAL\_RATE},
\texttt{PROPORTIONAL\_FAIR\_HIGH}, \texttt{PROPORTIONAL\_FAIR\_MEDIUM},
\texttt{PROPORTIONAL\_FAIR\_LOW}, \texttt{MAXIMUM\_C\_OVER\_I}. 
$R_{t+1}$ is the \textit{reward} received after taking \textit{action} $A_t$ given the state $S_t$. 
\note{It is important to note that in our case, the optimal configuration $A_t$ varies as state $S_t$ changes, unlike multi-armed bandit algorithms assuming an optimal action disregarding the system state \cite{shoaei2019reconfigurable,liao2019learning}. This will be illustrated in the results. 
}
In our experiment, we aimed to improve the throughput for the long-term. Thus we defined the reward as the spectrum efficiency. We also tried defining the reward as the gap between UEs' throughput, i.e the minimum UE throughput minus the maximum UE throughput. This design encouraged fairness between UEs. 
\note{
To make the training more stable, we clip the reward and KPIs to a small range to avoid gradient explosion. $\gamma$ is the discount factor. If it is larger, the return considers longer-term rewards. We fix it as 0.95. 
}
The goal of our model is to maximize the \textit{return} $G_t$ as follows,

\begin{equation}
G_t = \sum_{i=t}^{\infty} \gamma^{i-1} R_{i+1}(S_{i}, A_{i}), \quad 0<\gamma<1 
\end{equation}
\note{
A neural network to approximate the state-action value function, which is known as the Q-function. The size of the input layer is the same as the number of KPIs.
It has one hidden layer with 32 units with dense connections from the input layer and to the output layer. The output layer uses softmax, and the size is the same as the number of actions.
}
\begin{equation}
Q(s,a; \boldsymbol{\theta} ) 
\approx \max_{\pi} \mathbb{E} \; [G_t | S_t=s, A_t=a, \pi]
\end{equation}
Where $\pi$ is the policy, a mapping from states to action. With a good approximation of the Q-function, we select the optimal action given a state that maximizes the discounted sum of all future rewards. In each step, the action is taken according to $A_t =\arg\max_a Q(a,s;\boldsymbol\theta)$ with probability $1 - \varepsilon$ or is selected randomly with probability $\varepsilon$. $\varepsilon$ balances the trade-off between exploration and exploitation. In the beginning, $\varepsilon = 1$ is large to encourage the agent to explore the unknown environment. $\varepsilon$ decays exponentially down to 0.1 so that the operation relies more on the trained model. Recent researches propose many heuristic methods to efficiently approximate the Q-function, such as the double Q-learning, prioritized memory replay, multi-step TD learning, etc. \cite{hessel2018rainbow}. Here we implement the algorithm by combining several methods. The task is continuous instead of episodic, so there is no terminal state during the training. As in many offline learning algorithms \cite{mnih2015human,lillicrap2015continuous,hessel2018rainbow}, we train the model by replaying the experience. 
\note{
The replay experience buffer records 5,000 experience tuple $(S_t, A_t, R_{t+1}, S_{t+1})$ in time order. It is pre-loaded with historical data.
At each time step, the update rule for Q-function is the following. We randomly sample a batch of continuous segments of experience from the replay buffer since we are using n-step TD learning.
}
\begin{equation}
\boldsymbol{\theta}_{t+1} = \boldsymbol{\theta}_{t} + 
\alpha \left(
Y^{\text{DoubleQ}}_t -  Q(S_t, A_t; \boldsymbol{\theta}_t) \right)
\cdot \nabla_{\boldsymbol{\theta}} Q(S_t, A_t;\boldsymbol{\theta}_t)
\end{equation}
\begin{equation}
\begin{aligned}
Y^{\text{DoubleQ}}_t 
&:= R_{t+1} + \gamma R_{t+2} + ... \\
+& \gamma^{n-1} \tilde{Q} \left(S_{t+n}, 
\underset{a}{\arg\max} Q(S_{t+n},A_{t+n}; \boldsymbol{\theta}_t); \tilde{\boldsymbol{\theta}}_t \right)
\end{aligned}
\end{equation}
\begin{equation}
\tilde{\boldsymbol{\theta}}_{t+1} 
= (1-\tau) \tilde{\boldsymbol{\theta}}_{t}
+ \tau \boldsymbol{\theta}_{t+1}
\end{equation}
Two neural networks are used with parameters by $\boldsymbol{\theta}$ and $\tilde{\boldsymbol{\theta}}$ to alleviate the maximization bias or so called upward bias \cite{van2016deep,sutton2018reinforcement}. After one iteration, we partially update the parameter $\boldsymbol{\theta}$ into $\tilde{\boldsymbol{\theta}}$ to make the training more stable \cite{lillicrap2015continuous}. The computation for each iteration is much faster than the sampling time.

Working on a the real apparatus will bring more persuasive results. On the other hand, we have to compromise on the fine-tuning, as we can not fast forward the process in the real world, and the scalability. We utilize existing eNodeB reporting network measurements to derive KPIs. Cell level measurements are reported periodically like every minute, and UE level measurements are session-based. In our study, we build the KPI composer/publisher to aggregate KPIs once every minute.  
\section{Results} \label{sec:results}

\begin{figure}[ht]
\centering

\begin{subfigure}[h]{0.5\textwidth}
\centering
\includegraphics[width=1\linewidth]{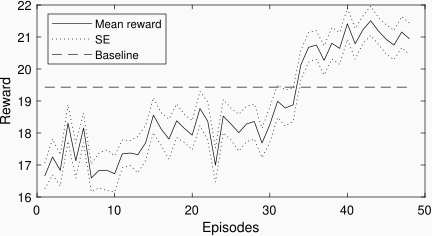}
\caption{Max throughput.}
\end{subfigure}
\begin{subfigure}[h]{0.5\textwidth}
\centering
\includegraphics[width=1\linewidth]{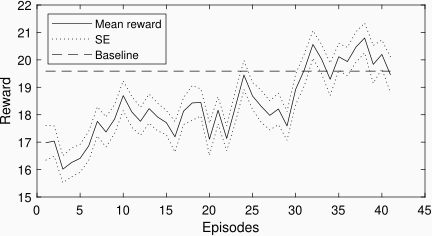}
\caption{Max throughput.}
\end{subfigure}
\begin{subfigure}[h]{0.5\textwidth}
\centering
\includegraphics[width=1\linewidth]{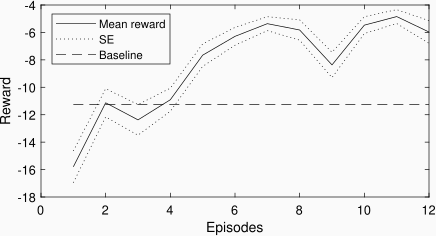}
\caption{Min UE gap.}
\end{subfigure}
\caption{Averaged reward of each episode. One episode is 90 minutes, including 90 steps. (a), (b) are experiments for optimizing the cell throughput, (c) is for optimizing the gap of UEs' throughput. The solid curve is the mean reward of each episode, and the dotted curve shows the standard error band of the mean. The dashed line represents the mean reward of all baseline episodes. } 
\label{fig:reward_results}
\end{figure}

Figure \ref{fig:reward_results} presents the results of 3 experiments. 
\note{
The first two are experiments for optimizing the cell throughput, where the reward is defined as the cell throughput. 
The third experiment is for optimizing the gap of UEs' throughput, in which the reward is defined as the difference between the minimum UE's throughput minus the maximal ones. 
The cell throughput and UE throughput are reported by the eNodeB directly.
}
The solid curve in Figure \ref{fig:reward_results} is the mean reward of each episode, and the dotted curve shows the standard error band of the mean. The standard error band is used to verify whether the improvement is significant or simply due to the variance of the data. \note{
The dashed line represents the mean reward of the baseline scenario, which is the best known constant action.
In the first two experiments, the action is set as \texttt{MAXIMUM\_C\_OVER\_I} as a constant with domain knowledge. In the third experiment, the action is set as \texttt{EQUAL\_RATE} as a constant.  Other worse constant actions are not shown.
}
In the beginning, the RL agent is not well trained, so the performance is below the baseline. As the training goes by, the performance elevates gradually and finally outperforms the baseline. The constant baseline policies can achieve acceptable performance. But to get further improvement, the model should take the dynamics of the system into the consideration, which is hard for a model designed with only domain knowledge.

\section{Discussion} \label{sec:discussion}

To improve the performance of the RL model and transfer the algorithm from the testbed to a massive network we need to consider the following inevitable challenges. 
\paragraph{Training and tuning}
Because the sampling rate of the platform was slow, we had to compromise on the thorough training and fine-tuning. This was the cost of dealing with the actual environment. Limited by these issues, we are not clear about the ceiling of the performance, but at least the model has the potential to do better than our results with longer experiments. 
In practice, algorithm debugging is another problem with this slow responding testbed.
So, we developed and tested the software framework in the OpenAI gym with coarse tuning to verify if the reward curve could boost up quickly in the first few episodes. We noticed the empirical clue that a model with a larger stable reward tends to have faster training in the first few episodes \cite{hessel2018rainbow}, so we pay more attention to the beginning sessions. Usually, a smaller model needs fewer data to train, so we chose simpler algorithms and smaller neural networks. But the model might not be rich enough to capture more complicated optimal policy. 
To get faster training, we tried prioritized experience replay \cite{schaul2015prioritized}. However, due to the noisy data, this method does not help too much. The large variance contaminated the weighted experience sampling, so it was hard to tell whether a sample had a large TD error representing or with large noise. Besides these efforts, we tried the pre-training method using historical records by following the method in \cite{hester2018deep}. Nevertheless, the historical data were not from an expert's demonstration, so the pre-training does not yield significantly better results. This was also seen in our OpenAI gym test experiments with only historical data but not experts' guidance. A promising method to train the model faster is to parallelize the process with more platforms running simultaneously and asynchronously, such as \cite{mnih2016asynchronous}. Except for these endeavors, we still do not have a good strategy for model tuning. We argue that this is not just an issue for our experimental design, but a limitation in general for reinforcement learning or optimal control dealing with the complicated real environment. 

\paragraph{Safety}
RL training is notorious for its instability that the reward curve can cliff fall unpredictably. If such a model is deployed in the RAN with millions of customers, the consequence can be catastrophic. This potential high risk leads to the conservativeness of the application in the industry. The RL agent is designed to maximize the reward in long term, so the dropping down of the reward in a few steps is not prohibited. In addition, the risk is related to the uncertainty of the environment, so even an optimal policy can perform poorly in some situations. Many exploration methods by default are blind to these risks of actions. On the other hand, without exploring the real environment, the model will hardly get a chance to exploit better. This concern has been addressed in the \textit{safety reinforcement learning} \cite{garcia2015comprehensive}. For our case, the strategies can be: 1. Train the model first then transfer it to the production without changing. 2. Gradually exploit the model from regional to global. 3. Prohibit several exploration options based on domain knowledge or historical records.

\paragraph{Customization}
While the reinforcement learning framework is very general for complicated environments, applying it to a wide range of problems still requires a deep understanding of the domain. A critical assumption of RL is that the underlying Markov decision process does not change or changes very slowly. The behavior of the testbed may be very different from that in the real network due to the spatial and temporal heterogeneity. So, patching the issues still requires further understanding and deep domain knowledge.

\paragraph{Combination of configurations}
In section \ref{sec:experiment} paragraph eNodeB, we list many aspects of the configurations for the eNodeB, but we only provide experiments for the MAC scheduler. It is attractive to combine several configurations to achieve multiple goals. 
A natural question is that if the RL agents were trained separately, for example for MAC scheduler and power on/off, and the two sub-agents share some parameters, how to merge them into a new one with multiple goals. This saves a lot of time from re-training the model from scratch. Another problem is that if the selected action from the two agents is different, then how to solve the conflicts. 

\note{
\paragraph{Scalability}
To isolate our testbed, all apparatus were kept in a Faraday room. 
To ensure reproducibility, each mobile device was controlled by a laptop to follow specific activities/traffic cycle. Some devices were covered by Faraday bags to weaken the RF signals to mimic real network RF variability. Due to these high maintenance requirements, our experiments started with small scale.  
}

\section{Conclusion} \label{sec:conclusion}
In this work, we first set up a RAN testbed in a lab environment for SON optimal policy design. Then we exploited recent advances of artificial intelligence and implement a deep reinforcement learning algorithm on the real environment and get satisfactory improvement for the complicated dynamic RAN. After stepping out of the academic RL research which heavily relies on simulators, we find more challenges such as lack of training data, model tuning, model scale-up, etc. Our work serves as a stepping stone for the following studies dealing with the real environment.


%





\ifCLASSOPTIONcaptionsoff
  \newpage
\fi





\bibliographystyle{IEEEtran}
\bibliography{IEEEabrv,Bibliography}
%


\vfill


\end{document}